\documentclass[11pt]{article}
\usepackage{amssymb,citesort}
\begin{document}

\title{The low temperature interface between
the gas and solid phases of hard spheres with a
short-ranged attraction}

\author{Richard P. Sear\\
~\\
Department of Physics, University of Surrey\\
Guildford, Surrey GU2 5XH\\
United Kingdom\\
email: r.sear@surrey.ac.uk}

\date{\today}

\maketitle

\begin{abstract}
At low temperature, spheres with a very short-ranged
attraction exist as a near-close-packed solid
coexisting with an almost infinitely dilute gas.
We find that the ratio of the interfacial tension between these two phases
to the thermal energy
diverges as the range of the attraction tends to zero.
The large tensions when the interparticle attractions are short-ranged
may be why globular proteins only crystallise over a narrow range
of conditions.
\end{abstract}



\section{Introduction}

The interfacial tension is a useful quantity to know, it not only
defines the cost of an interface but is also a central feature
in the classical nucleation theory of first order phase
transitions \cite{gunton83}. However, the interfacial tension
between solid and fluid phases is unknown for all but a very few
off-lattice microscopic models. Even for the very simple hard-sphere
potential a consensus on its value has only recently been reached
\cite{ohnesorge94,kyrlidis95}.
This is despite the fact that
we have known the bulk phase diagram of hard spheres for thirty
years \cite{hoover68}. The reason for the lack of calculations
of surface tensions is due to their difficulty:
calculation of the interfacial tension of hard spheres
is a formidable problem in density functional theory
\cite{ohnesorge94,kyrlidis95}. Here we calculate the interfacial
tension of hard spheres with a very short-ranged attraction.
The limiting case when the range tends to zero is the
$\beta_0$ model of Stell \cite{stell91,hemmer90}.
The phase behaviour of the $\beta_0$ model is straightforward, if a little
peculiar.  Above a certain temperature $T_{coll}$ the
behaviour is identical to that of hard spheres
(below close packing \cite{bolhuis94}) and below
this temperature they phase separate into an infinitely dilute gas
coexisting with a close-packed solid
\cite{stell91,hemmer90,bolhuis94,borstnik97,sear98}.

The bulk solid phase of spheres with a very short-ranged attraction
can be described,
due to its very high density, accurately and simply using
a cell theory \cite{buehler51}.
We will extend our previous cell theory treatment of the bulk
\cite{sear98} to the interface in order to calculate the
interfacial tension analytically.
Of course, the interfacial
tension between a solid and another phase depends on the
orientation of the interface with respect to the lattice of the
solid. We calculate it when the surface of the solid is one of a
few of the low index lattice planes.

The $\beta_0$ limit with its zero ranged
attraction is a purely mathematical limit,
unobtainable in an experiment. However, there are two types of
colloidal systems where the attraction has a range which is small
in comparison to that of the hard repulsive interaction.
These are mixtures of colloidal particles with either
smaller particles \cite{duijneveldt93,kaplan94,dinsmore95,imhof95}
(which may be surfactant micelles \cite{bibette91}) or small
polymer coils \cite{ilett95},
and globular protein molecules
\cite{broide91,broide96,rosenbaum96,muschol97}.

\section{Model and bulk phase behaviour}

First, we define the well-known square-well potential.
It is the spherically symmetric pair potential $u(r)$ defined by
\begin{equation}
u(r)=
\left\{
\begin{array}{ll}
\infty & ~~~~~~ r \le \sigma\\
-\epsilon & ~~~~~~ \sigma<r\le \sigma(1+\delta)\\
0 & ~~~~~~ r > \sigma(1+\delta)\\
\end{array}\right. ,
\label{monoss}
\end{equation}
where $\sigma$ is the hard-sphere diameter, and
$r$ is the separation between the centres of the spheres.
Here we will always be considering short-ranged attractions,
$\delta\ll1$. The first person to consider very short-ranged
attractions was Baxter \cite{baxter68} who considered a potential
with zero range, $\delta=0$, and with a well-depth $\epsilon/T$
adjusted so that the second virial coefficient was of order
unity. This model is often termed the sticky-sphere model.
Within it the second virial coefficient is used as
a temperature like variable. However, Stell \cite{stell91}
showed that the sticky-sphere model was pathological, its fluid 
phase was unstable at all non-zero densities.
Therefore, we will not consider this model but instead will
follow Stell when we take the
limit $\delta\rightarrow 0$,
thus obtaining his $\beta_0$ model \cite{stell91}.

The bulk phase behaviour of the $\beta_0$ model is described in Refs.
\cite{bolhuis94,stell91,hemmer90,sear98}.
If the two limits $\delta\rightarrow0$ and $T/\epsilon\rightarrow0$
are taken such that $T>T_{coll}$ then
the equilibrium phase behaviour is identical
to that of hard spheres. If $T<T_{coll}$ then
the behaviour is radically different: a close-packed
solid coexists with a fluid phase of zero density.
This close-packed solid may be either face-centred cubic or
hexagonal close-packed; both have the same number of nearest neighbours
and the same maximum density and so will have very similar
free energies. Because of this we do not specify which one is
actually formed. The temperature $T_{coll}$ is
\cite{hemmer90,stell91,sear98}
\begin{equation}
\frac{T_{coll}}{\epsilon}=
\frac{2}{\ln(1/\delta)}.
\label{tcoll}
\end{equation}
Note that we have used energy units for the temperature,
i.e., units in which Boltzmann's constant equals unity.

\section{Interfacial tension}

Above $T_{coll}$ the interface
between the coexisting fluid and solid phases is identical to that
of hard spheres; see
Refs. \cite{ohnesorge94,kyrlidis95,davidchack98} for work on the
interface between solid and fluid phases of hard spheres.
Below $T_{coll}$ the gas-solid interface will be very different. It will
be very narrow.
The free energies per particle ($/T$) in the coexisting solid and gas phases
approach $-\infty$ as the $\beta_0$ limit is taken, while
at intermediate densities, all densities which
are non-zero and below close packing, the free energy
is much higher \cite{sear98}.
Thus we expect the density will remain almost at its value in the bulk
solid even in the outermost layer of the solid before dropping
abruptly to zero. We will calculate the surface tension on the basis
that only the outermost layer of the solid differs from the bulk.
The solid layer beneath it, and the gas phase right up to the solid, are
assumed to be identical to the bulk solid and gas phases, respectively.

The free energy of a bulk solid phase can be estimated
using a cell theory \cite{buehler51}.
A cell theory starts from the 1-particle
partition function $q_1$ of a particle trapped in a cell
formed from its neighbouring
particles fixed at the positions they occupy in a ideal lattice.
The free energy
per particle $a_s$ is then obtained from
\begin{equation}
\frac{a_s}{T}=-\ln q_1,
\label{adef}
\end{equation}
where $q_1$ is defined to be in units of $\sigma^3$, and a term
which is the logarithm of the thermal volume of a sphere divided
by $\sigma^3$ is neglected.
For a close-packed solid $q_1$ is, see Ref. \cite{sear98},
\begin{equation}
q_1\simeq \delta^3\exp(6\epsilon/T) ~~~~~~~~
\rho>\frac{\rho_{cp}}{(1+\delta/2)^3}.
\label{q1}
\end{equation}
$\rho_{cp}=\sqrt{2}\sigma^{-3}$ is the close-packed density
of a close-packed solid of hard spheres.
The restriction on the minimum density of the solid ensures that
the sphere is close enough to all twelve of its neighbours to interact
via the attraction.
Inserting Eq. (\ref{q1}) into Eq. (\ref{adef}) yields
\begin{equation}
\frac{a_s}{T}=-3\ln\delta - \frac{6\epsilon}{T} ~~~~~~~~
\rho>\frac{\rho_{cp}}{(1+\delta/2)^3}.
\label{as}
\end{equation}
The first term in Eq. (\ref{as}) is the logarithm of the volume
available to the centre of a sphere and the second term is (half) its
energy of interaction with its twelve neighbours.
The volume available to the centre of a sphere is not precisely
$\delta^3$ but is $c\delta^3$, where $c$ is a prefactor of order
unity. We have neglected the $\ln c$ term in Eq. (\ref{as}) as
it is of order unity whilst the other terms diverge in the $\beta_0$ limit.

Equation (\ref{as}) gives the free energy per sphere in the interior
of the solid phase.
The free energy per sphere in the outermost layer
will be different. It will be given by an expression of the form
of Eq. (\ref{adef}) but in which $q_1$ is replaced by
the partition function of a particle in the outermost layer, $q_1^s$.
A particle in the outermost layer has fewer neighbours than in the bulk;
recall that the coexisting gas is at very low density
so there are almost no
interactions between the outermost layer and the gas.
How many fewer neighbours depends on which lattice plane
forms the outermost layer. We denote the number of missing neighbours
by $z_m$, it is equal to three for an interface in the 111 plane of
a face-centred-cubic lattice.
So, the energy of a particle
in the outermost layer is $(6-z_m/2)\epsilon$.
All we now need is the available volume for a particle
in the outermost layer.
The particle can only explore a volume greater than $\sim\delta^3$
at a cost of no longer being within $\delta$ of all its remaining
$12-z_m$ neighbours and so increasing the energy.
It is easy to see that if the particle moves over a distance
much larger than $\delta$ in any one direction then it can only
remain within $\delta$ of two particles: its motion consists
of rolling over the surfaces of a pair of adjacent spheres and
so is restricted to a volume of order $\delta^2\sigma$.
Thus, the entropy gain ($\times T$) is $-T\ln\delta$ but the energy cost
is $(10-z_m)\epsilon$. For the solid to be stable
the temperature must be below $T_{coll}$, Eq. (\ref{tcoll}), and
so $-T\ln\delta<2\epsilon$. The entropy gain is then is only greater than
the energy cost when $z_m>8$.
Similar arguments apply for allowing a particle
to move over a distance much larger than $\delta$ in two
or three directions. This is only favourable when
$z_m>7$ and $z_m>6$, respectively.
For any flat outer layer $z_m$ will be less than six,
and the outermost particles
will rattle inside a volume of order $\delta^3$ as they do in the bulk.
This is consistent with our assumption that only the
outermost layer of the solid differs from the bulk.
The particles in the layer below the outermost layer interact with
twelve neighbours and so have the same free energy as in the bulk,
apart from corrections of order $T$.

So, for $q_1^s$ we have
\begin{equation}
q_1^s\simeq \delta^3\exp\left[(6-z_m/2)\epsilon/T\right] ~~~~~~~~
\rho>\frac{\rho_{cp}}{(1+\delta/2)^3},
\label{q1s}
\end{equation}
which gives a free energy difference per particle
between the outermost layer and the interior of the solid of
\begin{equation}
\frac{a_i-a_s}{T}=\frac{z_m\epsilon}{2T} ~~~~~~~~
\rho>\frac{\rho_{cp}}{(1+\delta/2)^3}.
\label{ai}
\end{equation}
This difference can be converted into a surface tension by dividing
by the area per sphere in the outermost layer.
(In doing so we are implicitly fixing the surface of tension
to be that which fixes the surface excess number of particles
to be zero \cite{rowlinson82}.)
For example, for a 111
surface the area per sphere is $\sqrt{}3/2\sigma^3$ and $z_m=3$, so
\begin{equation}
\gamma_{111}=\sqrt{}3\epsilon\sigma^{-2}\simeq1.73\epsilon\sigma^{-2}
~~~~~~~~~~T<T_{coll}.
\label{gamma111}
\end{equation}
Similarly, for the 110 and 100 surfaces of a face-centred-cubic lattice the areas per sphere
are $\sqrt{}2\sigma^2$ and
$\sigma^2$, respectively and the $z_m$'s are 5 and 4.
Thus, $\gamma_{110}=(5/(2\sqrt{}2))\epsilon\sigma^{-2}
\simeq 1.77\epsilon\sigma^{-2}$
and $\gamma_{100}=2\epsilon\sigma^{-2}$.
The 111 surface has the lowest surface tension because it has the
lowest ratio of number of missing bonds per surface sphere to
area per sphere.
The surface tensions are all of order $\epsilon\sigma^{-2}$.
In the $\beta_0$ limit the ratio $\epsilon/T$ is infinite,
recall that we are below $T_{coll}$.
Thus, the ratio of the surface tension (expressed using the sphere diameter
as a unit of length) to the thermal energy is infinite.
As the range of attraction becomes very small the interfacial
tension becomes very large.

The assumptions which underlie the derivation of Eq. (\ref{gamma111})
should be valid whenever $\epsilon/T\gg1$, $\delta\lesssim0.1$ and the fluid
phase is highly dilute.
The first restriction ensures that the interface is only one
layer thick, the second that Eqs. (\ref{q1}) and (\ref{q1s}) are valid, and
the third that the outer layer of solid does not interact to a significant
degree with the fluid phase.
Under these conditions, Eq. (\ref{gamma111}) is a good approximation for
the interfacial tension and this tension is high.

An interfacial tension of order $\epsilon\sigma^{-2}$ is not a surprise.
It is what we would obtain if we just approximated the surface
tension by the {\em energy} per unit area
needed to pull a block of solid apart
to create two new surfaces; see the book of
Israelachvili \cite{israelachvili} where he estimates
interfacial tensions using just such an approximation.
Note that Eq. (\ref{gamma111}) has no explicit dependence on
the range of the attraction $\delta$, and so is a rough estimate
of the low temperature interfacial tension even of
the Lennard-Jones potential. Then $\epsilon$ would be the well-depth
of the Lennard-Jones potential.


We now compare our results with the earlier work of Marr and Gast
\cite{marr93,marr95}. This work was conducted within the Percus-Yevick
(PY) approximation for sticky spheres \cite{baxter68}.
PY for the sticky-sphere model was shown by Stell
\cite{stell91,hemmer90,borstnik97}
to yield qualitatively incorrect results; it predicts
vapour-liquid equilibrium at a temperature of $\epsilon/(\ln(1/\delta))$.
This is below $T_{coll}$, Eq. (\ref{tcoll}), and so in fact
the fluid phase is unstable at all non-zero densities.
Thus the results of Marr and Gast \cite{marr93,marr95}
are for the interface between phases which do not exist.

\section{Discussion and Consequences}

We have calculated the low temperature
interfacial tension of hard spheres with a very short-ranged attraction
and found that its ratio to the thermal energy per unit area
is very large.
Low temperature means below $T_{coll}$, Eq. (\ref{tcoll}).
In the limit that the range of the attraction tends to zero,
$\delta\rightarrow0$, the $\beta_0$ limit \cite{stell91},
then this ratio diverges.
In this limit above $T_{coll}$ the attractive part of the interaction
has a negligible effect:
all the equilibrium properties,
including the interfacial tension are identical to those of
hard spheres. At $T_{coll}$,
the coexisting fluid and solid densities
change discontinuously to zero and close-packing ($\rho_{cp}$),
respectively \cite{stell91,hemmer90,sear98}, and the
interfacial tension jumps to that given by Eq. (\ref{gamma111}).
If the range $\delta$ of the attraction is very small but finite
then the discontinuity at $T_{coll}$ becomes a
narrow temperature range \cite{note} over which the fluid and solid
densities at coexistence rapidly decrease and increase, respectively.
Below this temperature range the density of the fluid is very low
and our Eq. (\ref{gamma111}) for the interfacial tension
will be accurate.

When the range of the attraction is very small
there is only a narrow temperature range \cite{note} separating a
high-temperature regime
in which the spheres are almost hard spheres and a low-temperature
regime where they are almost at the low-temperature limit.
By almost at the low-temperature limit we mean that there is a very dilute
gas coexisting with a solid with a density near close packing, and
the interfacial tension between the two phases is then very high.
This may explain the finding of George and Wilson
\cite{george94} that there
is only a narrow slot in effective temperature within which globular proteins
can be made to crystallise \cite{broide96,rosenbaum96,muschol97}.
(Other, not necessarily contradictory,
explanations have been proposed by Poon \cite{poon97}, and
by ten Wolde and Frenkel \cite{tenwolde97}.)
The narrow slot may correspond
to the narrow temperature range where the coexisting densities and
interfacial tension are changing rapidly. Above this temperature
range the spheres are hard-sphere-like and so
only crystallise at high density,
above a volume fraction of 0.49 --- the density
at which a fluid of hard spheres coexists with the solid \cite{hoover68}.
At this density the dynamics of crystallisation may be slow
\cite{vanmegen91,underwood94} due to a nearby glass transition.
Below this temperature range the interfacial tension is very large.
The free energy
barrier to nucleation varies as the cube of the interfacial tension,
within classical nucleation theory \cite{gunton83}.
Thus, the rate of homogeneous nucleation varies as $\exp(-\gamma^3)$
and so is extremely small when the interfacial tension is large.
We conclude that spheres with a short-ranged attraction only
crystallise easily from a dilute solution
over a narrow temperature range around
$T_{coll}$: above it the spheres crystallise only at high density;
below it the interfacial tension is large and hence
homogeneous nucleation is extremely slow.





\begin{thebibliography}{99}
\bibitem{note} The temperature range when $\delta=0$, is zero
\cite{stell91,sear98}. For finite but short-ranged potentials it is small,
as can be seen in the computer
simulation results of Hagen and Frenkel \cite{hagen94}, and
in the theoretical results of Baus and co-workers
\cite{daanoun94,tejero95,coussaert95}.
Even as
the range of the potential increases towards the value at which a
liquid phase becomes stable then
the temperature range over which the coexisting fluid density
decreases from rather high values, volume fractions above 40\%, to
very low values is narrow.
This is enforced by the approach from below of
the metastable vapour-liquid critical point, to the
fluid--solid coexistence curve. Isotherms are
very flat near a critical point and so the approaching critical point
forces the curve of the coexisting fluid density (plotted
against temperature) to be very flat.
\end{thebibliography}
\end{document}